\documentclass[prl,twocolumn,superscriptaddress]{revtex4-1}
\usepackage{amssymb}
\usepackage{amsfonts}
\usepackage{bbm}
\usepackage{mathrsfs}
\usepackage{graphics,graphicx,epsfig,bm,amsmath,amsthm,amssymb}
\usepackage{bm}
\usepackage{bbm}
\usepackage{longtable}
\usepackage{multirow}
\usepackage{array}
\usepackage{color}
\usepackage{cases}
\usepackage{booktabs }
\usepackage[usenames,dvipsnames]{xcolor}

\usepackage{xcolor}

\usepackage{float}
\usepackage[a4paper,colorlinks=true,
linkcolor=blue,citecolor=blue,
pdfauthor={ },
pdftitle={ },
pdfsubject={ },
pdfkeywords={ }]{hyperref}

\begin{document}
\title{Searching for an Exotic Spin-Dependent Interaction between Electrons at the Nanometer Scale with Molecular Rulers}
\author{Man Jiao}
\affiliation{Hefei National Laboratory for Physical Sciences at the Microscale and Department of Modern Physics, University of Science and Technology of China, Hefei 230026, China}
\affiliation{CAS Key Laboratory of Microscale Magnetic Resonance, University of Science and Technology of China, Hefei 230026, China}
\affiliation{Synergetic Innovation Center of Quantum Information and Quantum Physics, University of Science and Technology of China, Hefei 230026, China}

\author{Xing Rong}
\email{xrong@ustc.edu.cn}
\affiliation{Hefei National Laboratory for Physical Sciences at the Microscale and Department of Modern Physics, University of Science and Technology of China, Hefei 230026, China}
\affiliation{CAS Key Laboratory of Microscale Magnetic Resonance, University of Science and Technology of China, Hefei 230026, China}
\affiliation{Synergetic Innovation Center of Quantum Information and Quantum Physics, University of Science and Technology of China, Hefei 230026, China}

\author{Hang Liang}
\affiliation{Hefei National Laboratory for Physical Sciences at the Microscale and Department of Modern Physics, University of Science and Technology of China, Hefei 230026, China}
\affiliation{CAS Key Laboratory of Microscale Magnetic Resonance, University of Science and Technology of China, Hefei 230026, China}
\affiliation{Synergetic Innovation Center of Quantum Information and Quantum Physics, University of Science and Technology of China, Hefei 230026, China}

\author{Yi-Fu Cai}
\affiliation{CAS Key Laboratory for Research in Galaxies and Cosmology, Department of Astronomy, University of Science and Technology of China, Hefei 230026, China}
\affiliation{School of Astronomy and Space Science, University of Science and Technology of China, Hefei 230026, China}

\author{Jiangfeng Du}
\email{djf@ustc.edu.cn}
\affiliation{Hefei National Laboratory for Physical Sciences at the Microscale and Department of Modern Physics, University of Science and Technology of China, Hefei 230026, China}
\affiliation{CAS Key Laboratory of Microscale Magnetic Resonance, University of Science and Technology of China, Hefei 230026, China}
\affiliation{Synergetic Innovation Center of Quantum Information and Quantum Physics, University of Science and Technology of China, Hefei 230026, China}

\date{\today}

\begin{abstract}

We propose that a type of molecular rulers, which contains two electron spins with precisely adjustable distance by varying the length of the shape-persistent polymer chains, can be utilized to constrain the axial-vector mediated interaction between electron spins  at the nanometer scale.
With measurements of the coupling strengths between two electron spins within different molecular rulers, an improved laboratory bound of exotic dipole-dipole interaction between electrons is established within the force range from 3 to 220 nm.
The upper bound of the coupling  $g_A^e g_A^e/4\pi\hbar c $ at 200 nm is $|g_A^e g_A^e/4\pi\hbar c| \leq 4.9\times10^{-13}$, which is about 20 times more stringent than previous limits.

\end{abstract}

\maketitle
Light bosons such as pseudoscalar fields (axion and the axion-like particles (ALPs)\cite{weinberg1978new,moody1984new}) and axial-vector fields (paraphotons and extra Z bosons) are hypothetically expected to address mysteries of fundamental sciences, namely, the microscopic origins of dark matter and dark energy, the resolution to the strong CP issue in quantum chromodynamics, as well as the possible connection with the hierarchy problem\cite{feng2010dark}.
The exchange of such bosons may mediate exotic spin-dependent interactions between ordinary fermions, which enables laboratorial searches on new particles via possible detections of new interactions\cite{moody1984new,dobrescu2006spin,PhysRevA.99.022113,safronova2018search}.
The exotic spin-dependent interactions are characterized by the dimensionless coupling constants between new bosons and fermions and the force range associated with the reduced Compton wavelength $\lambda$ of mediating bosons of mass $m$ \cite{dobrescu2006spin,PhysRevA.99.022113}.
There are experiments searching for the axial-vector dipole-dipole interaction between polarized electrons \cite{dobrescu2006spin}, ranging from atomic scale to the radius of earth \cite{safronova2018search}.
For the force range shorter than a millimeter, there are experiments such as Nitrogen-Vacancy (NV) centers in diamonds\cite{PhysRevLett.121.080402}, trapped ions\cite{kotler2015constraints},  single-atom Electron Spin Resonance with Scanning Tunneling Microscope (ESR-STM) \cite{choi2017atomic,luo2017constraints} and atomic spectroscopy \cite{ficek2017constraints}, providing stringent upper limit of the exotic interaction.
%In previous experiments, the assembly of the interacting particles are all realized by physical techniques,
%such as electric and magnetic fields to capture charged particles, electron beam evaporation in ultrahigh vacuum to enable individual atoms adsorb in films and confocal microscope used to focus the NV center dozens of microns away from polarized electrons.
\begin{figure}
\centering
\includegraphics[width=1\columnwidth]{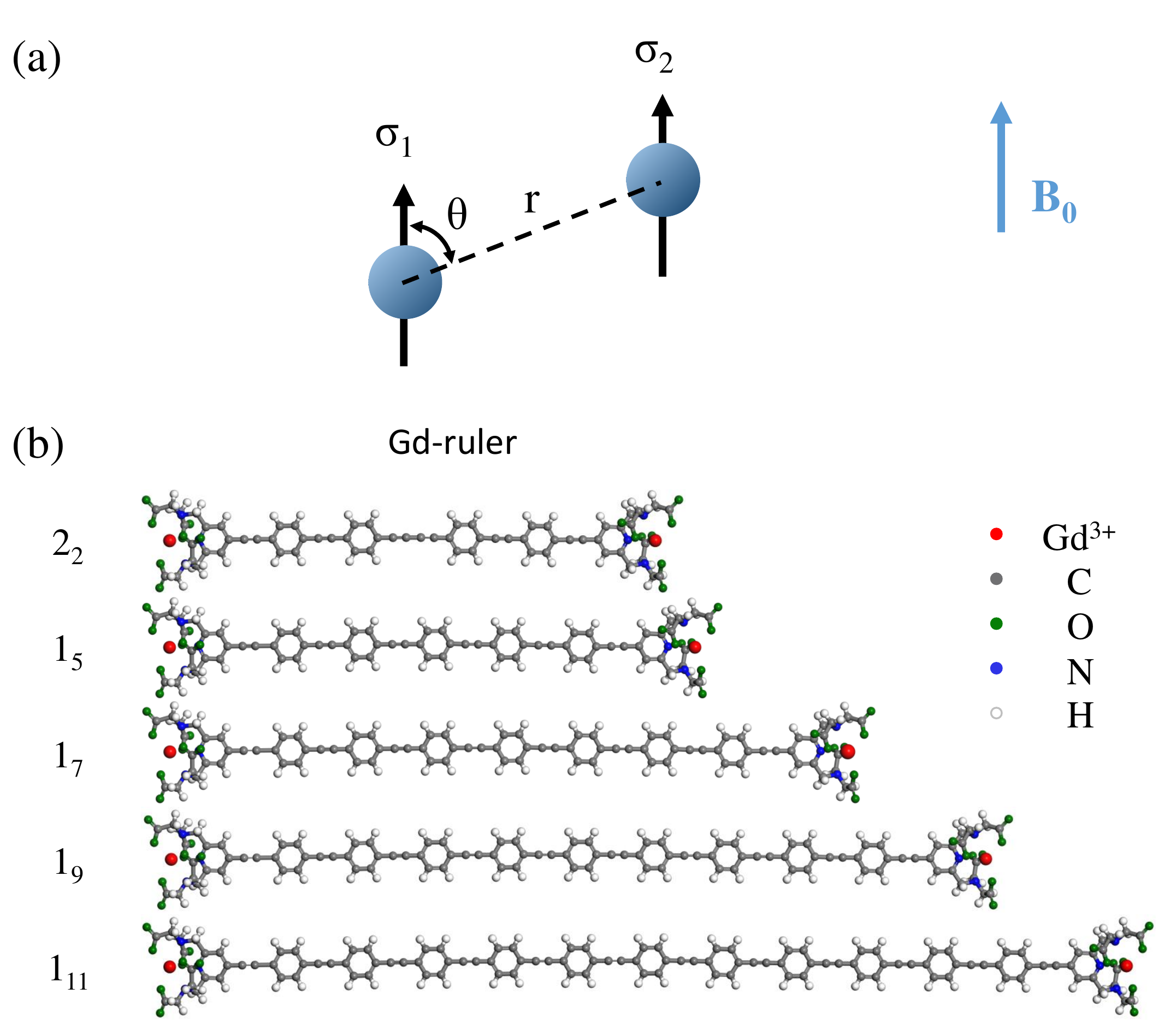}
\caption{ (a) A schematic diagram of two coupled electron spins ($\sigma_1$ and $\sigma_2$) in an external magnetic field $B_0$. The $\theta$ stands for the angle of the displacement vector between two spin labels, $\mathbf{r}$, and the external magnetic field, $\mathbf{B_0}$. (b) Structural formula of five molecular rulers with increasing length. The red dots denote the Gd atoms, the dark gray dots denote the carbon atoms, the green dots denote the oxygen atoms, the blue dots denote the nitrogen atoms and the dark gray circles denote the hydrogen atoms. Water-soluble side chains are omitted for clarity.
\label{figure1}}
\end{figure}

 In this paper, we propose a novel method to search for exotic spin-dependent interaction between electrons at the nanometer scale.
 A type of molecular rulers, which contains two electron spins, has been proposed to constrain the exotic spin-dependent interaction.
 The distance between two electron spins can be precisely set by varying the length of the shape-persistent polymer chains of the molecular rulers. The coupling strengths between the two electron spins can be obtained by Double Electron Electron Resonance (DEER) experiments.
 Then we established a  constraint on the axial-vector mediated interaction between electrons at the nanometer scale, which considerably improves on previous experimental bounds.

Figure \ref{figure1}a shows that two electron spins interact with each other in an external magnetic field. The magnetic dipole-dipole interaction between two electron spins is described by the following Hamiltonian:
\begin{equation}
H_d =  - \frac{\mu_0\gamma_1 \gamma_2\hbar^2}{16\pi r^3}[3(\vec{\sigma_1}\cdot\hat{r})(\vec{\sigma_2}\cdot\hat{r})- (\vec{\sigma_1}\cdot\vec{\sigma_2})],
\label{Hdd}
\end{equation}
where $\vec{\sigma_1}$ and $\vec{\sigma_2}$ stand for Pauli vectors of the two electron spins,
$\gamma_1,\gamma_1$ are gyromagnetic ratios of the two electron spins,
$r=|\vec{r}|$ and $\hat{r}=\vec{r}/r$ are the displacement and the unit displacement vector between two electron spins.
$\mu_0$ is the magnetic constant,
and $\hbar$ is Plank's constants divided by 2$\pi$.
The axial-vector dipole-dipole interaction mediated by hypothetical axial-vector bosons can be written as \cite{dobrescu2006spin},
\begin{equation}
H_2 = \frac{g_A^eg_A^e}{4\pi\hbar c}\frac{\hbar c}{r} (\vec{\sigma_1}\cdot \vec{\sigma_2})e^{{-\frac{r}{\lambda}}},
\label{H2}
\end{equation}
where $g_A^e g_A^e/4\pi\hbar c$ is dimensionless axial-vector coupling constant between electrons, $\lambda=\hbar/(mc)$ is the force range, $m$ is the mass of the hypothetical particle and $c$ is the speed of light.

To constrain the exotic interaction described by equation \ref{H2} within micrometer scale, several state-of-art methods  have been developed to place two electron spins with precise distance and to measure the coupling strength between them.
The coupling between two $^{88}S_r^+$ ions, each with a single electron spin, has been measured with high resolution \cite{Kotler_2014}, and the interaction can be constrained with micrometer scale \cite{kotler2015constraints}.
Single NV centers can be utilized as quantum sensors to measure the magnetic signal from polarized electrons, and the constraint was  improved for the force range from 10 to 900 $\mu$m \cite{PhysRevLett.121.080402}.
Recently, ESR-STM have been combined so that individual atoms can be precisely placed on a surface with atomic control \cite{Baumann417} and the coupling between them can be detected \cite{choi2017atomic}. With experimental data obtained from ESR-STM measurements on pairs of metal ions with tunable distances \cite{choi2017atomic}, constraint on the axial-vector dipole-dipole interaction at the nanometer scale can be established \cite{luo2017constraints}.

Our method is to measure the coupling between two electron spins located on two ends of a molecular ruler, and  to constrain the exotic dipole-dipole interaction at the nanometer scale.
Molecular rulers is a type of shape-persistent molecules, which contains several repeating units, such as p-phenylene (PP) and ethynylene (E).
The PP and E units have geometrically unambiguous structure and high stiffness \cite{jeschke2010flexibility}.
The lengths of molecular rulers can be precisely tuned by changing the number of repeating units.
A pair of spin labels with electron spins, can be attached on both ends of the molecular ruler.
Thus the distance between electron spins can be adjusted in the nanometer scale by varying the length of shape-persistent polymer chain.
DEER experiments can be utilized to measure the coupling between the electron spins.

A type of molecular rulers connecting two $ Gd_{(\rm \uppercase\expandafter{\romannumeral3})}$ labels (Gd-rulers) have been chosen as the system for constraining the exotic interaction.
$ Gd_{(\rm \uppercase\expandafter{\romannumeral3})}$ complex is a type of spin label with half integral high-spin(S=7/2), while the electron spin is in a $^{8}S$ ground state with 7 unpaired electrons in its f orbital \cite{dalaloyan2015gd}.
The chemically synthesized Gd-spacer-Gd compounds have a rodlike moiety consisting of PP and E units, acting as spacers which connect the two $ Gd_{(\rm \uppercase\expandafter{\romannumeral3})}$ spin labels as shown in Figure \ref{figure1}b.
The synthesis of these geometrically well-defined Gd-rulers is reported\cite{qi2016spacers}.
Precise DEER measurements of a series of Gd-rulers with different Gd-Gd distances are reported \cite{dalaloyan2015gd}.
As shown in Figure \ref{figure1}b, five Gd-rulers with lengths ranging from 4.25-8.78 nm are chosen to constraint the exotic interaction, and the $ Gd_{(\rm \uppercase\expandafter{\romannumeral3})}$-$ Gd_{(\rm \uppercase\expandafter{\romannumeral3})}$ pair can be treated as a pair of weakly coupled spin-half electron spins \cite{dalaloyan2015gd}. These molecules are denoted as Gd-rulers $\mathbf{1_{n}}$ (n = 5,7,9,11) and $\mathbf{2_2}$, which have different sequences of repeating units of PP and E as shown in Figure.\ref{figure1}b.

\begin{figure}[!h]\centering
\includegraphics[width=1\columnwidth]{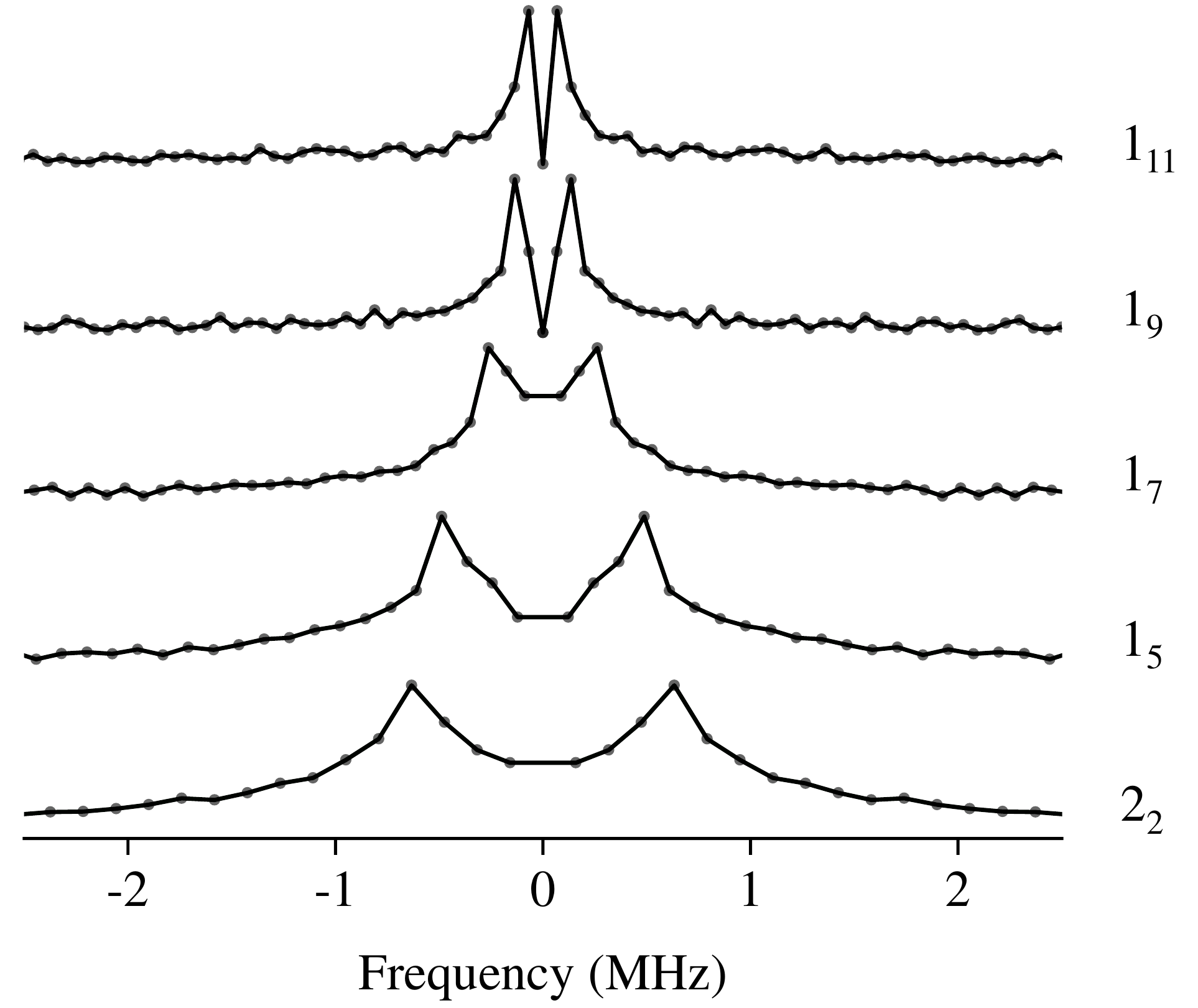}
\caption{
Experimental results of DEER measurements of Gd-rulers in frequency domain. Data are selected from Ref. \onlinecite{dalaloyan2015gd}.
%The splitting between two peaks of the spectra increases when the distance of two electron spins decreases.
The corresponding samples are labeled in the right side of the DEER spectra.
These Gd-rulers are synthesized in water soluble, and subjected to DEER measurements at W-band frequencies in shock-frozen solutions in a mixture of D$_2$O and glycerol-d$_8$ \cite{dalaloyan2015gd}.
}
\label{figure2}
\end{figure}

%and the dipolar pseudo-secular term and ZFS effect are take into consideration in systematic error discussed in constraints section, t
In the experiments, we consider the secular terms of the interaction Hamiltonians due to the large zeeman terms in the strong magnetic field.
The magnetic dipole-dipole interaction and the exotic interaction between two $Gd_{(III)}$ spin labels have the following form,
\begin{equation}
\left\{
\begin{aligned}
&H^{secular}_d =  - \frac{\mu_0\gamma_{Gd}^2\hbar^2}{16\pi r^3}(3\cos^2\theta-1)\sigma_{1}^z\sigma_{2}^z,\\
&H^{secular}_2 = \frac{g_A^eg_A^e}{4\pi\hbar c}\frac{\hbar c}{r}e^{{-\frac{r}{\lambda}}}\sigma_{1}^z\sigma_{2}^z.
\end{aligned}
\right.
\end{equation}
The magnetic dipole-dipole interaction $H^{secular}_d $ is dependent on the angle $\theta$ between external magnetic field and the displacement vector between of two spin labels in Gd-rulers.
Because the Gd-rulers are randomly oriented, the splitting due to the magnetic dipole-dipole interaction is the Pake patterns \cite{Pake_1948}.
If the additional splitting due to the possible $H^{secular}_2$ is taken into account, the strength of the coupling $\omega_{\perp}$ ($\theta = 90^\circ$) is half of the splitting of the two prominent features,
\begin{equation}
\omega_{\perp} =\frac{\mu_0\gamma_{Gd_{(\rm \uppercase\expandafter{\romannumeral3})}}^2\hbar^2}{4\pi h}\frac{1}{ r^3}+  \frac{g_A^eg_A^e}{4\pi\hbar c}\frac{4\hbar c}{h}\frac{e^{-\frac{r}{\lambda}}}{r}.
\label{omega_general}
\end{equation}
%The two prominent components of spectrum are the splitting due to total coupling when $\theta = 90^\circ$.
Figure \ref{figure2} shows the DEER results in frequency domain.
The value of the coupling, $\omega_{\perp}$, increases when the distance of two electron spins decreases.
The linewidths of the spectrum are partly due to the distributions of the distance, which are from the flexibility of the molecular building blocks\cite{jeschke2010flexibility,dalaloyan2015gd}.
The linewidths caused by flexibilities of the molecular rulers are less than one tenth of those of the experimental results \cite{SM}.
The decoherence of the electron spin labels also contributes to the experimental linedwiths.

%Figure \ref{figure3} shows the distance dependence of the splitting.
%The experimental results of intermolecular coupling between spin-labels are presented in Figure.\ref{figure2}.
%Coupling strength of the five molecules are presented as black squares, and the error bars are due to the frequency precision in Pake pattern.

\begin{figure}[ht]
\includegraphics[width=1.0\columnwidth]{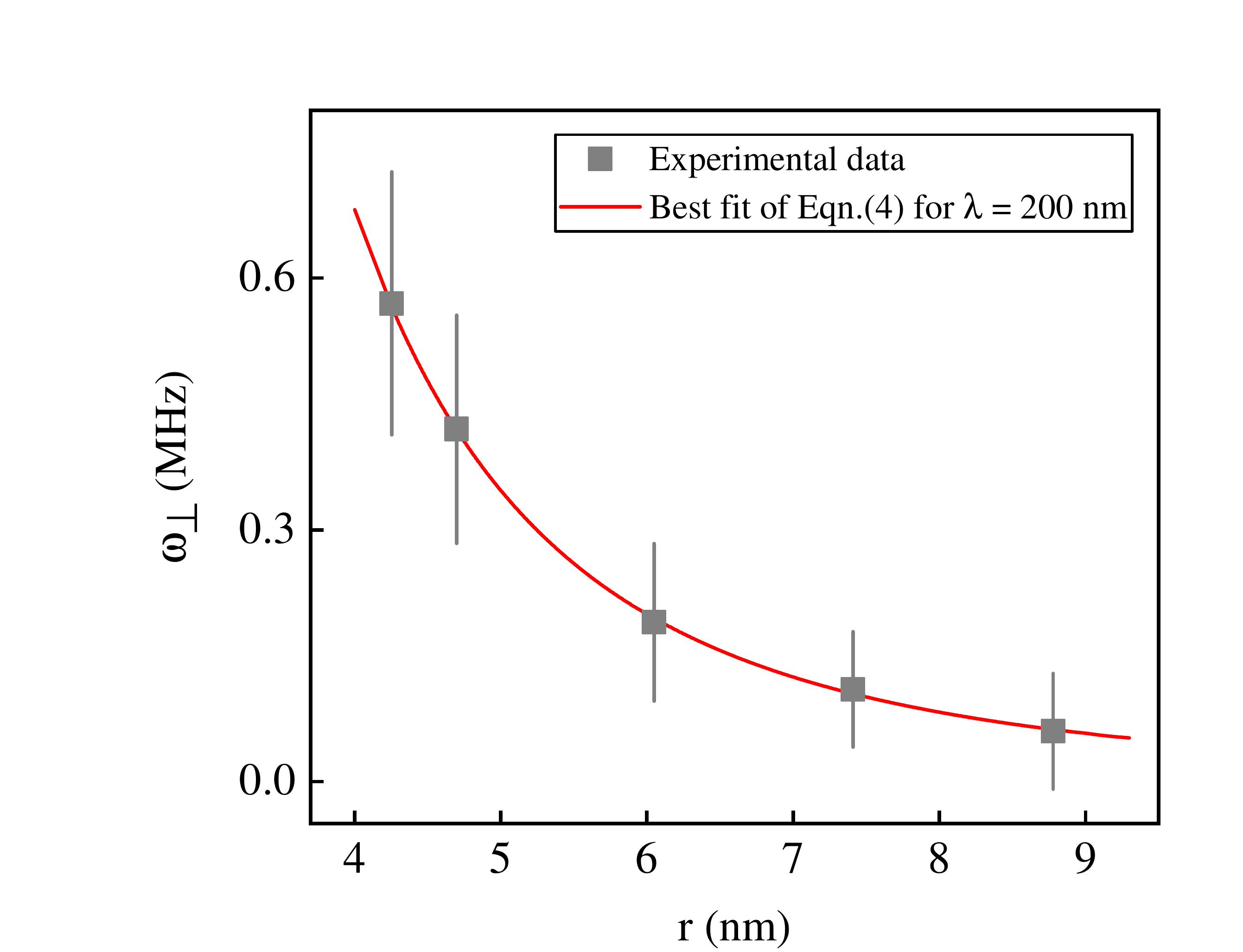}
\caption{
Measured couplings $\omega_{\perp}$ as a function of distances between two electron spins. The black squares with error bars are the experimental coupling strength between Gd-Gd spin-labels with different Gd-Gd distances. Error bars are obtained from the resolution of DEER spectrum which are essentially limited by the experimental linewidths. Red line is the fit for coupling strength with Equation \ref{omega_general} when $\lambda $= 200 nm.
 }
 \label{figure3}
 \end{figure}

%Assuming the intermolecular coupling frequency is contributed from both the magnetic dipole-dipole interaction and the axial-vector-mediated interaction, fittings for different values of the fixed force range $\lambda$ give constraints for the exotic interaction between electrons.
The experimental results and a fitting of the couplings, $\omega_{\bot}$, between spin-labels with distance from 4.25-8.78 nm are presented in Figure \ref{figure3}.
Black squares are the experimental values of the couplings $\omega_{\perp}$, which are half of the splitting of the peaks in Figure. \ref{figure2}.
Due to the geometrically unambiguous and high stiffness of the Gd-ruler's backbone at low temperatures, the distance between $ Gd_{(\rm \uppercase\expandafter{\romannumeral3})}$ can be estimated from density functional theory (DFT).
We used scalar relativistic DFT as implemented in the Gaussian software package\cite{g16} and performed full geometry optimizations for all structures.
In our DFT calculations, the three-parameter hybrid exchange functional of Becke and the exchange functional of Lee¨CYang¨CParr (B3LYP) \cite{PhysRevB.37.785,Becke1993,PhysRevA.38.3098} was employed.
More precisely, to ensure a correct approach for the open-shell species, we employed the spin-unrestricted open-shell version of this functional (UB3LYP). The standard basis set 6-31G(d) is used for C, H and N atoms while the Stuttgart RSC 1997 ECP basis set \cite{Andrae1990} is used for Gd atom.
The DFT results shows that Gd-Gd distance for Gd-rulers $\mathbf{2_2}$, $\mathbf{1_5}$, $\mathbf{1_7}$, $\mathbf{1_9}$ and $\mathbf{1_{11}}$, are $4.25$, $4.70$, $6.05$, $7.41$ and $8.78$ nm, respectively.
The red line in Figure \ref{figure3} is a fitting to the experimental data with Equation \ref{omega_general} when the force range $\lambda $= 200 nm.
The gyromagnetic ratio of the $ Gd_{(\rm \uppercase\expandafter{\romannumeral3})}$ spin obtained by this fit is $\gamma_{ Gd_{(\rm \uppercase\expandafter{\romannumeral3})}}=(0.92 \pm 0.01) \gamma_e$, where $\gamma_e = 2\pi\times2.8~$MHz/Gauss.
The coupling constant of the exotic axial-vector-mediated interaction for $\lambda = 200~$nm is obtained to be $g_A^eg_A^e/4\pi\hbar c =  (-1.42 \pm 1.80 )\times 10^{-13}$.
The value of the axial-vector field induced interaction is less than its errors showing no evidence of the exotic interaction observed in these measurements.
The constraint for other given force range $\lambda$ can be obtained with the same procedure.

%The constraint on the $g_A^eg_A^e$ can be derived with different values of $\lambda$.

%One systematic error is due to the uncertainty of Gd-Gd distance, the difference from DFT analysis and Tikhonov regularization on DEER measurement are considered as the uncertainty of Gd-Gd distance.
%For example, the Gd rulers $\mathbf{1_{11}}$ has an distance uncertainty of 0.28 nm, to estimate the correction of the uncertainty in  Gd-Gd distance to $\omega_d$, we randomly take $10^5$ samples of Gd-Gd distance in Gd ruler $1_{11}$ which satisfies a Gaussian distribution, the correction to  $\omega_d$ is  $0.0004\pm 0.006$ MHz, which is much smaller than the statistical accuracy of 0.07 MHz.
%The correction to $\omega_d$ can be obtained for the other Gd-rulers with the same procedure, and the correction is much smaller than the statistical accuracy, thus the systematic error can be neglected.
%%The systematic errors of Gd-Gd distance are much smaller than statistical accuracies .
%The presence of the dipolar pseudo-secular term and the ZFS are taken into account as systematic error and also proved to be much smaller than statistical accuracies.
%The correction of frequency domain spectrum of $ Gd_{(\rm \uppercase\expandafter{\romannumeral3})}$ pairs with ZFS interaction and dipolar pseudo-secular term are followed by Ref. \cite{dalaloyan2015gd}.
%Through perturbation theory and exact diagonalization, the correction of coupling frequency of Gd ruler $1_{11}$ is much smaller than the statistical accuracy of 0.07 MHz.

\begin{figure}[!h]\centering
\includegraphics[width=1.0\columnwidth]{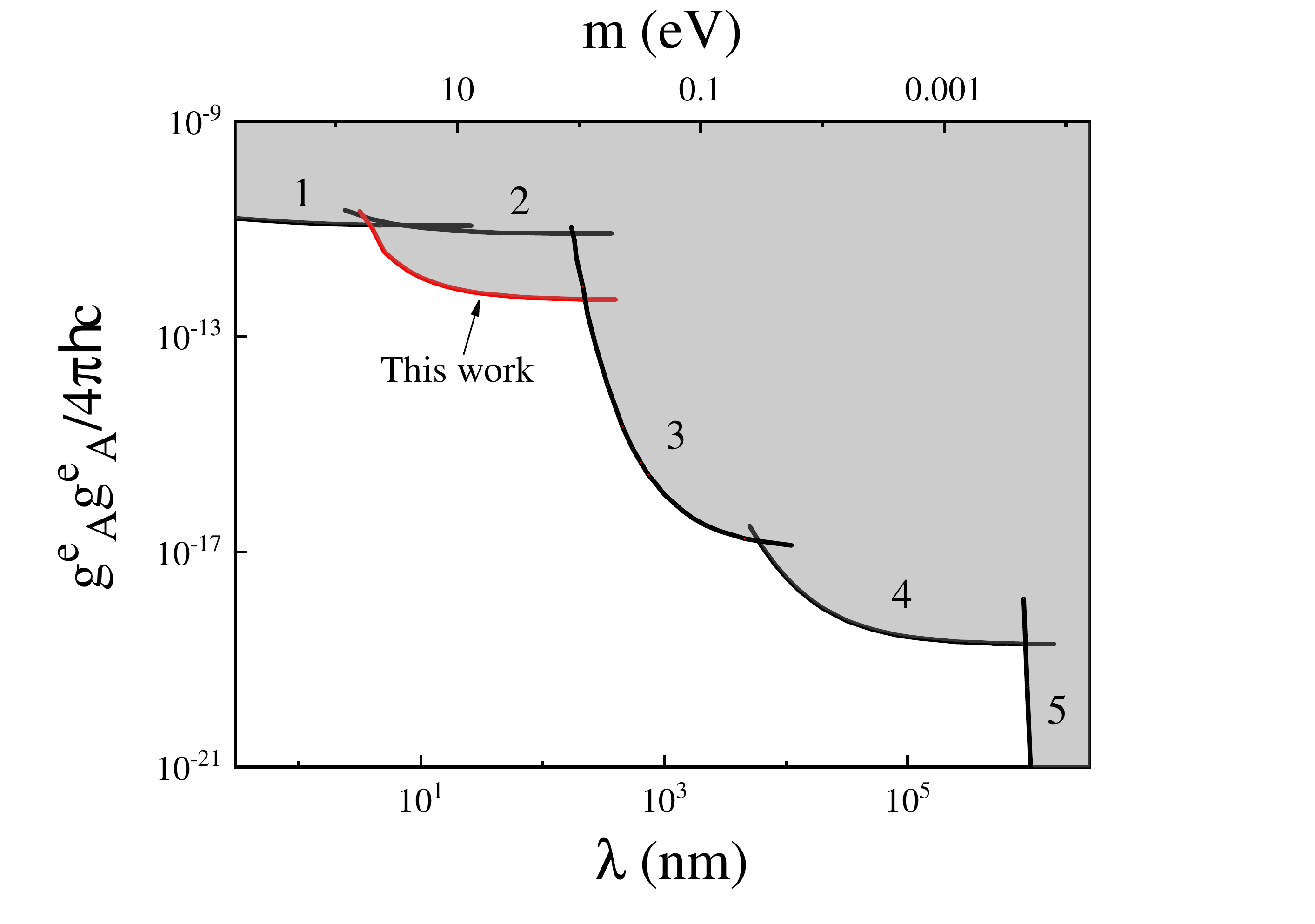}
\caption{\label{figure4} Upper limit on the axial-vector-mediated dipole-dipole interactions between electrons, $g_A^e g_A^e/4\pi\hbar c$, as a function of the force range, $\lambda$, and the mass of the axial-vector bosons, m.  The black solid lines represent upper bounds obtained from Refs.\cite{ficek2017constraints,PhysRevLett.121.080402, luo2017constraints,kotler2015constraints,Ritter:1990aa}. This work (the red line) establishes a new laboratory bound in the force range from 3 to 220 nm. The obtained upper bound of the interaction at 200 nm is  $|g_A^e g_A^e/4\pi\hbar c|\leq 4.9\times10^{-13}$, which is one order of magnitude more stringent than previous experiment.}
\end{figure}

Figure \ref{figure4} shows the new constraint obtained from this work together with recent constraints from experimental searches for axial-vector-mediated dipole-dipole interactions.
The horizontal axis shows the force range of the interaction, inversely proportional to the mass of the boson communicating the interaction.
The vertical axis shows the dimensionless coupling parameter $g_A^e g_A^e/4\pi\hbar c$ between electrons.
Filled areas correspond to excluded values.
For the force range 0.22$<\lambda<$  10 $\mu$m, the constraint was established by Kotler et al.\cite{kotler2015constraints}.
For the force range $\lambda <$ 3 nm, the upper limit was obtained from Ref.\cite{ficek2017constraints}.
For the force range from 10 to 200 nm, a previous limit was set by Luo et al.\cite{luo2017constraints} with data from ESR-STM experiments \cite{choi2017atomic}.
The linewidths of the ESR-STM experiments are about several Megahertz, which are due to the short coherence time of the ions.
Electron spin labels in molecular rulers have longer coherence times than those of metal ions in ESR-STM experiments.
The linewidths of the DEER experiments for Gd-rulers are about sub-Megahertz.
The method here can provide better precision of detection for coupling between two electrons compared with ESR-STM experiments.
The red line is the constraint established in this work, which clearly shows that more stringent constraints in the range from 3 to 220 nm.
Specifically, the obtained upper limit of the exotic dipole-dipole interaction at 200 nm is about a factor of 20 more stringent than the one obtained from Ref. \cite{luo2017constraints}.

%The constraints can be further improved by suppressing intermolecular background and extend the measurement times to get higher precision in frequency domain.
%DEER measurements are widely used for distance determination in biomolecules.
%In this paper,  benefit from the backbone made up from the PP and E units has geometrically unambiguous structure and high stiffness,
%Gd-Gd distance is obtained by DFT analysis, which is independent of DEER measurements.
%Assuming the DEER measurements resolve both magnetic dipole-dipole interaction in the Standard Model and the exotic interaction beyond the Standard Model between spin labels,
%fittings of measured DEER frequencies with two interactions with different r dependence yield stringent constraints on exotic spin-dependent interaction between electrons at nanometer scale.

In summary, we provide a novel method to search for an exotic spin-dependent interaction at the nanometer scale.
An improved constraint on axial-vector mediated interaction between electron spins has been established at the nanometer scale.
DEER measurements have been carried on molecule with triple spin labeling with $Gd^{3+}$, $Mn^{2+}$, and a nitroxide \cite{Wu_2017}, these measurements based on chemically synthesized molecules may provide new possibilities for searching for exotic spin-dependent  interactions.
We also expect that spin labels of molecular rulers with longer coherence times can be  synthesized for this study.
Since longer coherence times ensure longer observation times, better frequency resolutions can be achieved and the constraint can be further improved.

Authors thank Prof. Dieter Suter for helpful discussion. We are also grateful to Prof. Daniella Goldfarb for sharing us the experimental data of the DEER experiments. We do appreciate Xiaojing Liu for the DFT calculations. This work was supported by the National Key R$\&$D Program of China (Grants No. 2018YFA0306600 and No. 2016YFB0501603), the Chinese Academy of Sciences (Grants No. GJJSTD20170001, No.QYZDY-SSW-SLH004 and No.QYZDB-SSW-SLH005), and Anhui Initiative in Quantum Information Technologies (Grant No.
AHY050000). X.R. thanks the Youth Innovation Promotion Association of Chinese Academy of Sciences for support.
Y.F.C is supported in part by the NSFC (Nos. 11653002, 11722327), by National Youth Thousand Talents Program of China, by the CAST Young Elite Scientists Sponsorship Program (2016QNRC001), and by the Fundamental Research Funds for the Central Universities.

We noticed that a revision of spin-dependent interactions have recently been proposed \cite{PhysRevA.99.022113}. We provide the constraints on the revisited interactions in the Supplemental Material \cite{SM}.

\clearpage

\onecolumngrid
\vspace{1.5cm}
\begin{center}
%\textbf{\large Supplementary Material}
\textbf{\large Supplementary Material for\\~\\Searching for an Exotic Spin-Dependent Interaction between Electrons at the Nanometer Scale with Molecular Rulers}
\end{center}
\twocolumngrid
\setcounter{figure}{0}
\setcounter{equation}{0}
\setcounter{table}{0}
\makeatletter
\renewcommand{\thefigure}{S\arabic{figure}}
\renewcommand{\theequation}{S\arabic{equation}}
\renewcommand{\thetable}{S\arabic{table}}
\renewcommand{\bibnumfmt}[1]{[RefS#1]}
\renewcommand{\citenumfont}[1]{RefS#1}

%\begin{appendix}

%\section{appendix}

\section{ 1. Distance distribution of the molecular rulers}
\begin{figure}
\includegraphics[width=1\columnwidth]{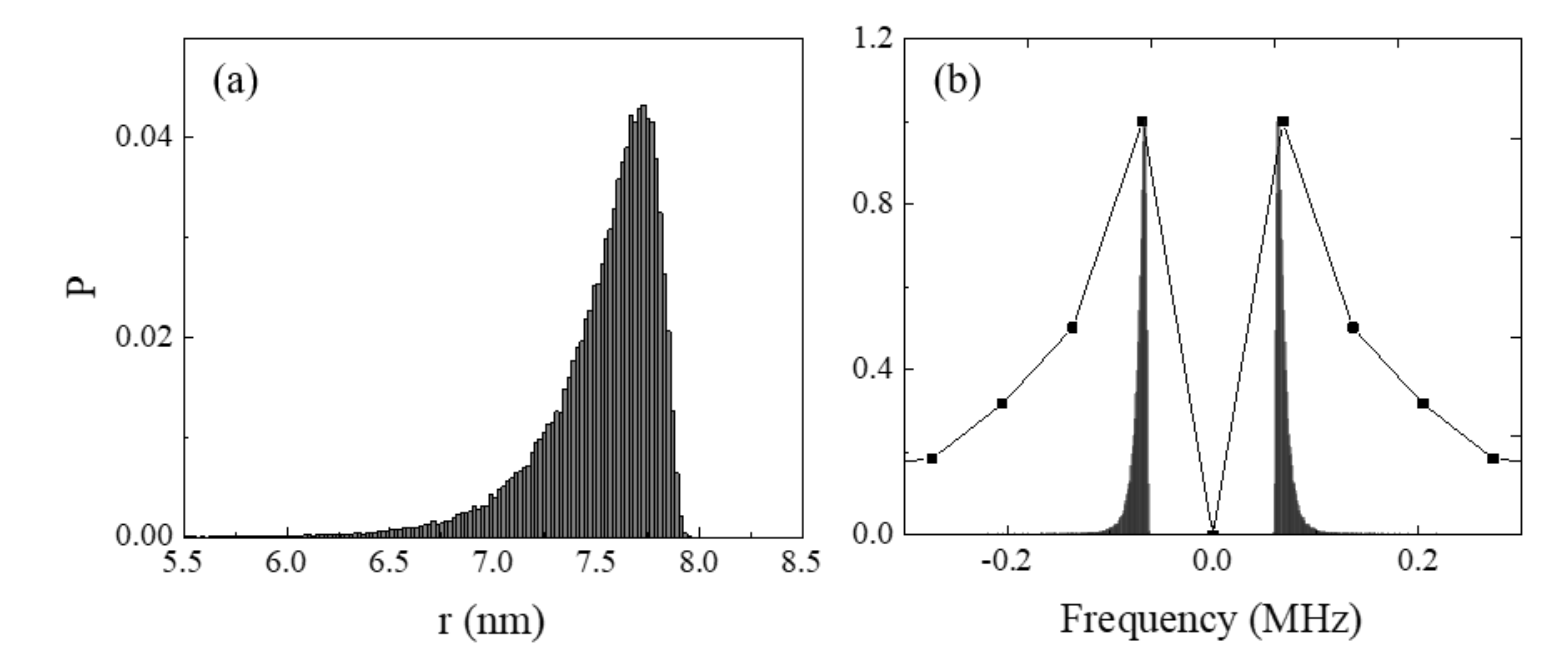}
\caption{(a) Histogram of distance distribution of the backbone of Gd-ruler $1_{11}$. (b) The Black square is the experimental result of Gd-ruler $1_{11}$ and the gray histogram is the distribution of the coupling strength due to the flexibility of the molecular ruler.}
\label{sfig1}
\end{figure}

%The molecular ruler containing several repeating units of p-phenylene (PP) and ethynylene (E) have geometrically unambiguous structure and high stiffness.
In ESR experiments where the signal is contributed from ensemble molecules, it's necessary to take the flexibility of  molecules into account.
The flexility of molecules can be estimated by the distance distribution of the molecular backbone.
The distance distribution of the backbone containing repeating units of p-phenylene (PP) and ethynylene (E) can be estimated by harmonic segmented chain model(HSC), where the distance distribution is determined by the bending potential F and the thermal energy $k_BT$ \cite{jeschke2010flexibilitySM}.
Here we take Gd-ruler $1_{11}$ as an example, which contains 11 repeating units of PP and E. The distance distribution derived from HSC model is shown in Figure \ref{sfig1}a with the FWHM of the distance distribution being $0.397 nm$.

The linewidths of DEER spectrum caused by flexibilities can be estimated by the first term of Eqn.4 in the main text.
As shown in Figure \ref{sfig1}b, the FWHM of the coupling strength caused by flexibilities is $0.008~$MHz, which is much smaller than that of the experimental result which is about $0.1054~$MHz.
It shows that the molecular rulers with high stiffness are good platform to investigate the spin-dependent interactions.

\section{2. Constraints on Revisited spin-dependent interactions}
\begin{figure}
\includegraphics[width=\columnwidth]{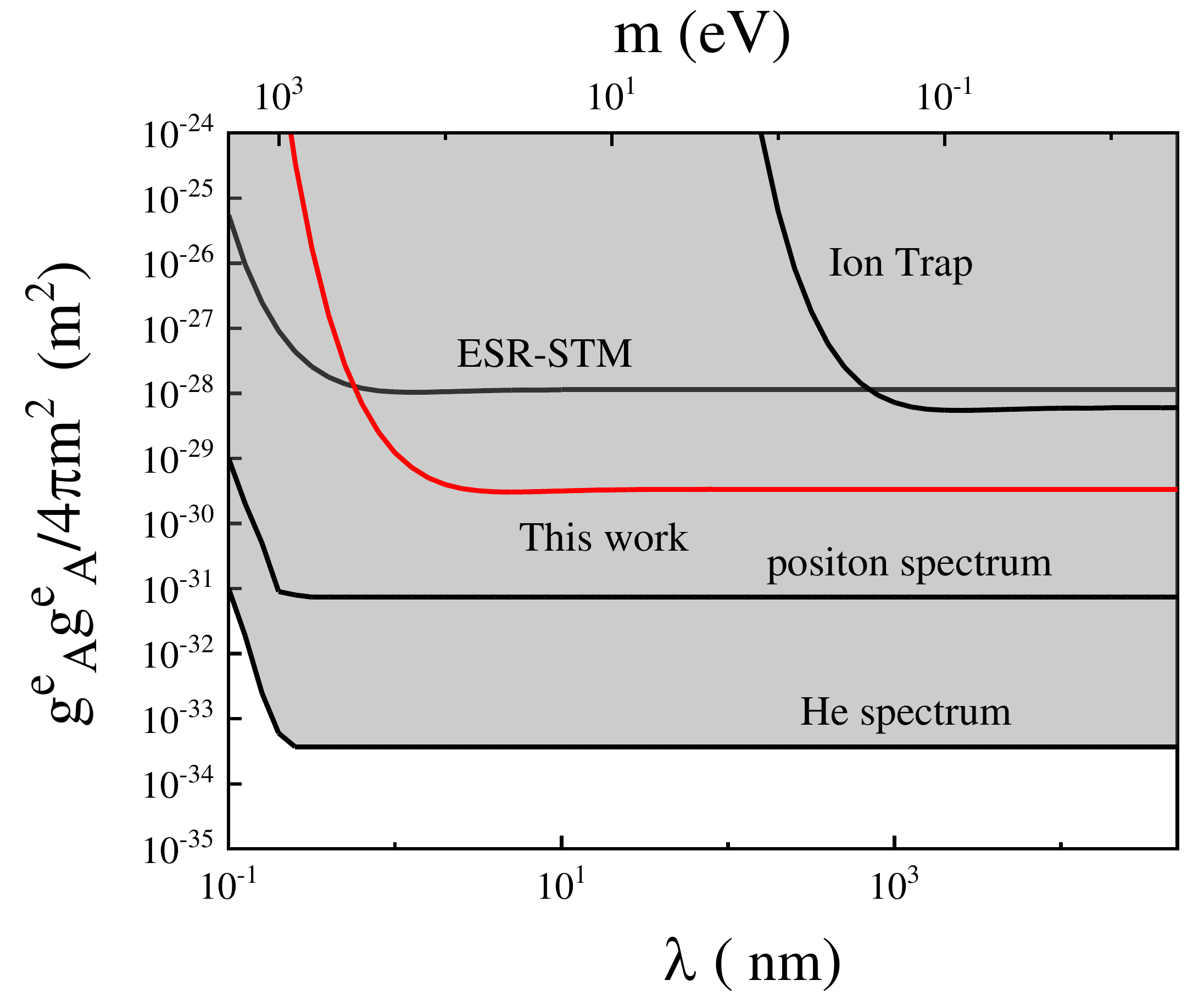}
\caption{ Upper limit on $|g_A^eg_A^e|/4\pi m^2$ between electrons, as a function of the force range and the mass of the axial-vector bosons, m. The black solid lines represent upper bounds obtained from Refs. \cite{luo2017constraintsSM,ficek2017constraintsSM,kotler2015constraintsSM,karshenboim2010precisionSM,leslie2014prospectsSM}. This work provides constraints as the red line.
\label{sfig2}
}
\end{figure}

Recently, a revision of spin-dependent interactions has been proposed, which gives nine non-relativistic potentials by the exchange of a spin-0 or spin-1 boson in coordinate-space \cite{PhysRevA.99.022113SM}.
Different from the spin-dependent interactions derived in a mixed momentum and coordinate-space proposed in 2006 \cite{dobrescu2006spinSM}, the revisited spin-dependent interactions in coordinate-space sort the spin-dependent interactions in a different way.
Our result can also provide constraints on the revisited spin-dependent interactions. We consider the axial-vector potential mediated by a massive spin-1 boson of the revisited interactions, the interaction between electrons can be written as \cite{PhysRevA.99.022113SM},
\begin{widetext}
\begin{equation}
\begin{array} { c } { V _ { A A } ( \boldsymbol { r } ) = - \frac { g _ { A } ^ { e } g _ {A } ^ { e } } { 4 \pi m ^ { 2 } } m ^ { 2 } \hbar c  \boldsymbol { \sigma } _ { 1 } \cdot \boldsymbol { \sigma } _ { 2 } \frac { e ^ { - m r } } { r } }  { - \frac { g _ { A } ^ { e } g _ { A } ^ { e } } { 4 \pi m ^ { 2 } } \hbar c \left[ \left( \boldsymbol { \sigma } _ { 1 } \cdot \boldsymbol { \sigma } _ { 2 } \right) \left( \frac { 1 } { r ^ { 3 } } + \frac { m } { r ^ { 2 } } + \frac { 4 \pi } { 3 } \delta ( r ) \right) \right. }{ - \left( \boldsymbol { \sigma } _ { 1 } \hat { \boldsymbol { r } } \right) \left( \boldsymbol { \sigma } _ { 2 } \hat { \boldsymbol { r } } \right) \left( \frac { 3 } { r ^ { 3 } } + \frac { 3 m } { r ^ { 2 } } + \frac { m ^ { 2 } } { r } \right) ] e ^ { - m r } } \end{array}
\end{equation}
\end{widetext}
where $m$ is the mass of the hypothetical particle, $g_A^e g_A^e$ is dimensionless axial-vector coupling constant between electrons and $c$ is the speed of light.
We take the combination of parameters $|g_A^eg_A^e|/4\pi m^2$ as a constant and give constraints on it (Fig. \ref{sfig2}).

%shows the constraints on this revisited interaction together with the data from ESR-STM and Ion trap experiment.This work also provide constraints as shown in the red line.
%our data establishes an new laboratory bound in the force range larger than 1nm, and would give a 30 times more stringent constraint at 200 nm.
%atomic spectroscopy\cite{ficek2017constraints} is not discussed.

\end{document}